\documentstyle[epsfig]{mn}

\newif\ifAMStwofonts

\ifoldfss
  \ifCUPmtlplainloaded \else
    \NewTextAlphabet{textbfit} {cmbxti10} {}
    \NewTextAlphabet{textbfss} {cmssbx10} {}
    \NewMathAlphabet{mathbfit} {cmbxti10} {} 
    \NewMathAlphabet{mathbfss} {cmssbx10} {} 
  \fi
  \ifAMStwofonts
    \ifCUPmtlplainloaded \else
      \NewSymbolFont{upmath} {eurm10}
      \NewSymbolFont{AMSa} {msam10}
      \NewMathSymbol{\upi}     {0}{upmath}{19}
      \NewMathSymbol{\umu}     {0}{upmath}{16}
      \NewMathSymbol{\upartial}{0}{upmath}{40}
      \NewMathSymbol{\leqslant}{3}{AMSa}{36}
      \NewMathSymbol{\geqslant}{3}{AMSa}{3E}

       \let\ge=\geqslant
    \fi
  \fi
\fi 

\ifnfssone
  \newmathalphabet{\mathit}
  \addtoversion{normal}{\mathit}{cmr}{m}{it}
  \addtoversion{bold}{\mathit}{cmr}{bx}{it}
  \newmathalphabet{\mathbfit} 
  \addtoversion{normal}{\mathbfit}{cmr}{bx}{it}
  \addtoversion{bold}{\mathbfit}{cmr}{bx}{it}
  \newmathalphabet{\mathbfss} 
  \addtoversion{normal}{\mathbfss}{cmss}{bx}{n}
  \addtoversion{bold}{\mathbfss}{cmss}{bx}{n}
  \ifAMStwofonts
    \ifCUPmtlplainloaded \else
      \UseAMStwoboldmath
      \makeatletter
      \new@mathgroup\upmath@group
      \define@mathgroup\mv@normal\upmath@group{eur}{m}{n}
      \define@mathgroup\mv@bold\upmath@group{eur}{b}{n}
      \edef\UPM{\hexnumber\upmath@group}
      \new@mathgroup\amsa@group
      \define@mathgroup\mv@normal\amsa@group{msa}{m}{n}
      \define@mathgroup\mv@bold\amsa@group{msa}{m}{n}
      \edef\AMSa{\hexnumber\amsa@group}
      \makeatother
      \mathchardef\upi="0\UPM19
      \mathchardef\umu="0\UPM16
      \mathchardef\upartial="0\UPM40
      \mathchardef\leqslant="3\AMSa36
      \mathchardef\geqslant="3\AMSa3E

       \let\ge=\geqslant
    \fi
  \fi
\fi 

\ifnfsstwo
  \DeclareMathAlphabet{\mathbfit}{OT1}{cmr}{bx}{it}
  \SetMathAlphabet\mathbfit{bold}{OT1}{cmr}{bx}{it}
  \DeclareMathAlphabet{\mathbfss}{OT1}{cmss}{bx}{n}
  \SetMathAlphabet\mathbfss{bold}{OT1}{cmss}{bx}{n}
  \ifAMStwofonts
    \ifCUPmtlplainloaded \else
      \DeclareSymbolFont{UPM}{U}{eur}{m}{n}
      \SetSymbolFont{UPM}{bold}{U}{eur}{b}{n}
      \DeclareSymbolFont{AMSa}{U}{msa}{m}{n}
      \DeclareMathSymbol{\upi}{0}{UPM}{"19}
      \DeclareMathSymbol{\umu}{0}{UPM}{"16}
      \DeclareMathSymbol{\upartial}{0}{UPM}{"40}
      \DeclareMathSymbol{\leqslant}{3}{AMSa}{"36}
      \DeclareMathSymbol{\geqslant}{3}{AMSa}{"3E}

       \let\ge=\geqslant
    \fi
  \fi
\fi 

\ifCUPmtlplainloaded \else
  \ifAMStwofonts \else 
    \def\upi{\pi}
    \def\umu{\mu}
    \def\upartial{\partial}
  \fi
\fi

\title{Pre-enriched, not primordial ellipticals}
\author[A. E. Sansom \& R. N. Proctor]
       {A. E. Sansom, R. N. Proctor \\
       Centre for Astrophysics, University of Central Lancashire,
Preston PR1 2HE}
\date{Accepted 1999 December 15.
      Received 1999 December 14;
      in original form 1999 October 11}
\pubyear{1999}

\begin{document}

\maketitle

\begin{abstract}
We follow the chemical evolution of a galaxy through star formation
and its feedback into the inter-stellar medium (ISM), starting from 
primordial gas and allowing for gas to inflow into the region being modelled. 
We attempt to reproduce observed spectral line-strenghts for early-type 
galaxies 
to constrain their star formation histories (SFH). The efficiencies and 
times of star formation are varied as well as the amount and duration of
inflow. We evaluate the chemical enrichment 
and the mass of stars made with time. Single stellar population (SSP) 
data are then used to predict line-strengths for composite stellar 
populations. The results are compared with observed line-strengths in ten
ellipticals, including some features which help to break the problem of
age-metallicity degeneracy in old stellar populations. We find that the
elliptical galaxies modelled require high metallicity SSPs ($>3$ Z$_{\odot}$)
at later times. In addition the strong lines observed cannot 
be produced by an initial starburst in primordial gas, even if a large 
amount of inflow is allowed for during the first few $\times 10^8$ years. 
This is because some pre-enrichment is required for lines 
in the bulk of the stars to approach the observed line-strengths in
ellipticals. These strong lines are better modelled by a system with 
a delayed burst of star formation, following an early SFH which can be 
a burst or more steady star formation. Such a model is representative 
of star formation in normal ellipticals or spirals respectively, 
followed by a starburst and gas inflow during a merger or 
strong interaction with a gas-rich galaxy. Alternatively,
a single initial burst of normal stars with a Salpeter initial mass 
function could produce the observed strong lines if it followed
some pre-enrichment process which did not form long-lived stars 
(e.g. population III stars).

\end{abstract}


\section{Introduction}

Many previous studies of star formation histories
in galaxies assumed solar metallicity stars throughout (e.g. Bruzual \&
Charlot 1993; Fritze - v. Alvensleben \& Gerhard 1994 -- hereafter FG94).
Arimoto \& Yoshii (1986) were amongst the first to account for 
non-solar metallicity stars in their photometric and chemical 
evolution models. They showed that the instantaneous recycling
approximation is invalid for stellar populations that have 
light from long-lived, metal-poor giant stars. More recent models include 
better coverage of the different stages of stellar evolution as well 
as stars with different metallicities (Vazdekis et al.\ 1996 -- hereafter
V96; Greggio 1997). V96 explored the effects of varying the initial stellar
mass function (IMF) and were able to generate colours and some of the 
line-strength indices in three early-type galaxies. 

In this paper we build self-consistent chemical evolution models
for different SFH in early-type galaxies, using published single stellar
populations. We allow for gas inflow but assume a Salpeter IMF for the 
distribution of stellar masses. These models are aimed at helping us to
better understand the strengths and gradients of spectral features
(lines, bands and colours) observed in early-type galaxies.
Tantalo et al.\ (1996) allowed for inflow and outflow, but modelled 
only colours, not line-strengths. They showed that inflows can solve the
problem of too many metal poor stars produced in closed box models
which use a single IMF. We attempt to fit a range of spectral features defined
by Worthey et al.\ (1994), some of which are sensitive to metallicity, 
others to age and yet others to both age and metallicity. This will 
provide a more stringent test of models than previous fits to galaxy
colours and broad spectral features such as the Mg$_2$ index which are all
degenerate to age and metallicity effects (Worthey 1994 -- hereafter W94). 
This degeneracy is such that populations of young, metal rich stars look 
like populations of old, metal poor stars.
 
A population of stars made at a single age and with a single metallicity
is known as a single stellar population (SSP). These can provide the
building blocks for evaluating the evolution of spectral features in
composite stellar populations made up of stars with a range of ages and
metallicities. We use SSPs from W94 to generate predicted feature strengths
for different star formation histories. W94 emphasised
that real galaxies contain stars covering a range of metallicities and 
probably a range of ages as well. So it is necessary to build a composite
model in which the star formation and enrichment are followed with time 
in order to
predict accurate spectral features seen now.  The purpose of building
such models is to improve our understanding of the SFH in galaxies,
through use of the detailed observations which can be obtained from
nearby galaxies. This approach complements, and must be consistent with,
the coarser information which can be obtained from more distant galaxies.

This paper is set out as follows: Section 2 discusses the
sources of enrichment data for supernovae (SN) and stars as well as
the parameters and values used in our chemical evolution models. In Section 3
the model input, methods and output are described and compared to other
models. Comparisons of best fit models to observed data for 10 ellipticals
are shown in Section 4. Discussion (including allowed model parameter
space and effects of varying input parameters) is given in Section 5,
where a brief discussion of trends with velocity dispersion is also
included. Conclusions are given in Section 6.

\section[]{Stellar and Gaseous Inputs}

The first step in our modelling is the choice of input data for the 
feedback from stars to the ISM. This 
includes the choice of star and SN data. Many results for yields and
masses are taken from tables of published data, linearly interpolated
for the appropriate metallicity $Z$ (mass fraction of elements heavier than 
helium) and star mass in our models. We make liberal use of linear 
interpolation to try to cover varying masses, metallicities and ages.
However we do not extrapolate stellar data but take the nearest case when
extremes of these variable spaces are reached in the models. Keywords for
input parameters are given in capital letters in the text and all the
input parameters are listed in table 1 which is discussed further in
Section 3.

\begin{table*}
 \centering
 \begin{minipage}{140mm}
  \caption{Input parameters and their values for (A) the 'primordial' and
(B) the 'merger' (delayed starburst) reference models. Initial mass of gas 
is $10^6$M$_{\odot}$ in these models. Parameters in bold font are the ones 
searched for best fits to ellipticals in Section 4.}
  \begin{tabular}{llll}
 &&& \\
PARAMETER   &\multicolumn{2}{c}{PARAMETER VALUES} &PARAMETER \\
NAME        & (A)     & (B)      &DESCRIPTION AND UNITS. \\
 &&& \\          
        &'Primordial' &'Merging'  & \\
         &Elliptical  &Spirals  & \\
{\bf ROOC0}      &0.7       &0.2    
&Initial constant $C$ (Gyr$^{-1}$) in SFR$=C\rho^{\alpha}$ \\
AL          &1.0       &       &Index $\alpha$ in SFR equation \\
RCRIT       &0.0       &       &Critical mass density for SF 
(M$_{\odot}$ per volume) \\
FLOSSLIM    &0.1       &       &Fraction for significant star mass loss \\
SNIA\_RATE  &$3.8E-5$  &       &SNIa (Number M$_{\odot}^{-1}$ Gyr$^{-1}$) \\
DT          &0.1       &       &Time step (Gyr) \\
FLOWRATE0   &$1.0E+7$  &0.0    &Initial inflow rate (M$_{\odot}$ Gyr$^{-1}$) \\
SNH         &70.0      &       &Upper star mass limit (M$_{\odot}$) \\
AM          &1.35      &       &Slope for IMF (1.35=Salpeter IMF) \\
TYPIMF      &S         &       &S=Single slope IMF \\
SSPDATA     &W         &       &SSP data (W=Worthey 1994) \\
X0          &0.7718    &       &Initial H mass fraction in gas \\
Y0          &0.2280    &       &Initial He mass fraction in gas \\
Z0          &0.0002    &       &Initial metal mass fraction in gas \\
{\bf TCHANGE1}    &0.3       &12.0   &1st discrete time change (Gyr)   \\
{\bf ROOC1}       &0.7       &4.0    &1st changed SFR constant (Gyr$^{-1}$) \\
{\bf FLOWRATE1}   &0.0       &1.0E+7 &1st changed inflow rate  
(M$_{\odot}$ Gyr$^{-1}$) \\
{\bf TCHANGE2}    &18.0      &12.1   &2nd discrete time change (Gyr)   \\
ROOC2       &0.7       &4.0    &2nd changed SFR constant (Gyr$^{-1}$) 
(=ROOC1) \\
FLOWRATE2   &0.0       &       &2nd changed inflow rate  
(M$_{\odot}$ Gyr$^{-1}$) \\
RICH        &Y         &       &Enriched inflow (Y=yes or N=no) \\
BHMASS      &6.0       &       &Mass of CO core for BH formation
(M$_{\odot}$) \\
TIME        &17.0      &       &Age ago when stars first started forming
(Gyr) \\
 &&& \\
\end{tabular}
\end{minipage}
\end{table*}

\subsection{SNII}
Woosley and Weaver (1995 -- hereafter WW) produced comprehensive 
models of enrichment by supernova explosions of massive stars. They give
ejected masses of individual elements as well as overall ejected mass.
Since element ratios are proving to be a useful diagnostic in
understanding the SFH in ellipticals (e.g. Davies, Sadler \& Peletier 1993
-- hereafter DSP93) we use these data (from WW) for stars 
between 10 and 40 solar masses (M$_{\odot}$) (the upper limit of their 
models). We assume their case A for stars up to 25 M$_{\odot}$ and their 
case B for more massive stars, following the assumptions made by Timmes,
Woosley \& Weaver (1995) in their study of our 
Galaxy. Case B corresponds to higher energies and more mass being blown 
off during the SN, whereas case A (from WW) for massive stars corresponds 
to more mass collapsing back into a black hole. This is effected 
by use of a piston mechanism acting at different layers, with different
energies for different cases of their models. Drawbacks of the WW 
models are:

\begin{enumerate}
  \item They do not allow for any stellar mass loss in a wind prior to
the SN explosion.
  \item They assume stars more massive than 40 M$_{\odot}$ contribute
nothing to the chemical enrichment of the ISM and future stars made from
it in a galaxy.
\end{enumerate}

Both the above points are increasingly poor assumptions for more metal rich 
and massive stars (Maeder 1992 -- hereafter M92). M92 modelled the yields 
from a wider range of initial star masses and, more importantly, modelled mass
loss in winds prior to any explosion. Such mass loss can explain helium-rich, 
massive star supernovae (SNIb) and mass loss observed from massive, main 
sequence stars (Wolf-Rayets), singly or in binaries. Table 2 shows the 
range of initial masses ($M_i$) and metallicities for which the WW or M92 
models of massive stars are more appropriate, assuming the fractional mass 
lost in winds (FLOSSLIM) is considered significant for fractions $>$ 0.0, 
0.1 or 0.9.
FLOSSLIM=0 corresponds to using only M92 data, whereas FLOSSLIM=1.0 
corresponds to using only WW data (where available, plus M92 data for 
$>40$ M$_{\odot}$). FLOSSLIM=0.1 is the compromise we use in most of our 
models and corresponds to taking M92 data for stars in which the M92 
star models indicate $> 10$ per cent mass is lost in a wind and taking 
WW data (where available) for less significant mass loss.
We use this compromise because a drawback of the M92 data is that it
only predicts yields for a few elements and overall metallicities,
whereas the WW data gives yields for many individual elements of interest.
 
The parameter FLOSSLIM also allows us to test the dependance of our models
on the uncertainties in the SN models and their yields.
In particular we predict abundances of Mg and Fe in our models, and for 
the M92 models we have to make some assumption about the yields of these
elements. We assume their yields are the same fractions of the 
metal yield as for stars from WW with the same initial mass. This is 
only an estimate since models of stars with heavy mass loss are likely 
to produce different fractional yields of Mg and Fe than models with 
no mass loss. This is about the best we can do until more complete 
SNII yields are available for a wider range of initial masses, and for 
models incorporating mass loss and its yields via stellar winds prior 
to any explosion. Metal rich ($\sim$ Z$_{\odot}$), massive stars 
(40 M$_{\odot}$ or greater) throw large amounts of helium 
and light metals (C and O) back into the ISM during their wind
phase and in some cases so much mass is lost that the remnant becomes a
neutron star or white dwarf rather than a black hole (M92).

\begin{table*}
 \centering
 \begin{minipage}{140mm}
  \caption{Table showing where different SNII models apply for different
assumptions about the level of significant mass loss. If mass loss is
ignored WW data are used (where available), if mass loss is considered 
significant M92
data are used. M92 data are used for high masses since there are no 
WW models for masses above 40 M$_{\odot}$. A blank means that M92 models
predict $<10$ per cent mass loss prior to any explosion for that 
metallicity and initial mass. An x corresponds to regions with
$>10$ per cent mass loss and an {\bf X} corresponds to regions with
$>90$ per cent mass loss from M92. This shows that the high mass and/or high 
metallicity stars are most affected by mass loss in winds prior to any
SNII explosion (M92).}  
  \begin{tabular}{l|c|c|c|c|c|c|c|c|c|c|c}
 Metallicity    & \multicolumn{11}{c}{ $M_i$  Initial masses (M$_{\odot}$)} \\
 $Z_i$ (Z$_{\odot}$) &12 &13 &15 &18 &20 &22 &25 &30 &35 & 40 & $>40$  \\
   $0.0$        &   &   &   &   &   &   &   &   &   &    & x \\
   $10^{-4}$    &   &   &   &   &   &   &   &   &   &    & x \\
   $0.01$       &   &   &   &   &   &   &   &   &   &    & x \\
   $0.1$        &   &   &   &   &   &   &   &   &   &    & x \\
   $1.0$        &   &   & x & x & x & x & x & x & x & {\bf X} & {\bf X}\\
   $>1.0$       &   &   & x & x & x & x & x & x & x & {\bf X} & {\bf X}\\ 
\end{tabular}
\end{minipage}
\end{table*}

M92 also considered several cases corresponding to different amounts of
mass trapped into a black hole. In their models this depends on the CO
core mass left just prior to the SN explosion. Their case A corresponds to
no mass going into black holes. This produces low helium-to-metal mass
increase ($\Delta Y/\Delta Z$) above primordial compared with observed 
values of $\Delta Y/\Delta Z$ between 3 and 6 from observations of 
extragalactic \hbox{H\,{\sc ii}} regions (Pagel et al.\ 1992). More 
realistically their case C (M92) corresponds to
stars with CO cores $>6$ M$_{\odot}$ all collapsing back into black holes,
thus trapping some heavy elements. We allow this core mass limit to be 
an input parameter (BHMASS) in our models. We use BHMASS=6 M$_{\odot}$
in most of our models.

Our other SNII input parameters are: the upper mass for stars (taken to be
$M_i = 70$ M$_{\odot}$ in most of our models); the lower mass limit for SNII 
(for which we assume $M_i = 10$ M$_{\odot}$) and the shape of the IMF. 
We assume an IMF consistent with that used to generate the SSPs used later
in our models (e.g. Salpeter's law with slope 1.35 for the SSPs of W94).
Richer and Fahlman (1996) recently searched for evidence of variations in the 
IMFs for different stellar environments and metallicities, but could find
no evidence for variable IMFs. On the other hand several authors have
suggested that starburst galaxies show evidence for a different IMF 
with a low mass cut-off of $>3$ M$_{\odot}$ (Wright et al.\ 1988). Such a
top heavy IMF may also help explain the quantities of metal enrichment in
clusters (Elbaz et al.\ 1995). We briefly discuss the effects of different
assumptions about the IMF in Section 5.
The main sequence lifetime for
a 10 M$_{\odot}$ star is about $3\times 10^7$ years, so this is the
shortest time step we can use in our models which assume the SNIIs
evolve within the time step in which they were made.

\subsection{SNIa}
Supernovae in binaries contribute to the heavy element enrichment in galaxies,
and occur throughout most of the history of the galaxy. We use the well
known standard model of Nomoto et al.\ (1984) (their W7 model) which 
predicts complete disruption of a CO white dwarf. A thermonuclear 
front converts most of the white dwarf to iron plus a few other heavy 
elements. SNIa return no H or He in these models, so they only contribute 
to increasing the metallicity with time in the ISM and subsequently formed 
stars. This has implications for the helium-to-metal mass increase with
time in the ISM, which may provide a constraint on the validity of any 
chemical evolution model.

SNIa occur in binary systems which take about a Gyr to evolve to the SN
stage (e.g. Timmes et al.\ 1995). Thus a time delay of 1 Gyr is included 
in our models before stars contribute to ISM enrichment through SNIa 
explosions. The SNIa rate per total mass of stars present 1 Gyr ago is one 
of our input parameters (SNIA\_RATE). This rate is assumed constant with time. 
The SNIa mass is thus taken from stars made a Gyr ago unless there is
not enough star mass available then, in which case the SNIa mass is taken 
from stars made at earlier times, allowing for some lower mass stars to 
evolve into SNIas. The nominal
rate in most of our models is taken from the rate inferred by
Timmes et al.\ (1995) for our Galaxy (0.53 century$^{-1}$), allowing for 
a Galaxy mass of $1.4 \times 10^{11}$ M$_{\odot}$. This gives
a rate of $3.8 \times 10^{-5}$ SNIa Gyr$^{-1}$ M$_{\odot}^{-1}$. 
Cappellero et al.\ (1993) give SN numbers in different galaxy types.
For SNIa+Ib/c they find $\sim 0.15$ SN per 100 years for E/S0 galaxies 
and $\sim 0.4$ SN per 100 years for Sab to Sc galaxies (averaging their
samples from three different catalogues of galaxies). Their data indicate 
that SNIb/c rates are $\sim 25$ to 30 per cent of the SNIa rates.
Taking a typical galaxy mass as $\sim 10^{11}$ M$_{\odot}$ this 
gives a SNIA\_RATE of between 1.5 and 4 $\times 10^{-5}$ Gyr$^{-1}$
M$_{\odot}^{-1}$, including SNIa+Ib/c. Our nominal rate is within 
this range. The effects of varying this rate are discussed in Section 5.

\subsection{Intermediate mass stars (IMS)}

The yields from IMS are taken from Renzini and Voli (1981).
IMS contribute mainly C,N,O plus H and He via stellar winds and planetary
nebulae. They are another source of delayed enrichment. We take the mass
range for IMS to cover $0.8<M_i<10$ M$_{\odot}$. We
assume they contribute most of their mass to the ISM after their main
sequence (MS) lifetimes. So there is a time delay before they contribute 
to the ISM enrichment. The timescales to evolve to white dwarfs are taken 
from Wood (1992) and the mass locked up in remnants is taken from the 
formula given in Wood (1992), and estimated from M92 (their figure 4)
for low Z. These parameters vary as a function of initial mass and 
metallicity at the time the stars formed. 

\subsection{Low mass stars}

For stars with $M_i<0.8$ M$_{\odot}$ we assume that they contribute 
negligibly to the chemical enrichment of the ISM. They remain on 
the main sequence
and loose very little mass throughout their evolution to date.
We fix the lower limit to the mass range in our models at 
0.1 M$_{\odot}$ to be consistent with Worthey's SSPs, and we assume 
that lower mass stars or brown dwarfs play no part in the chemical 
evolution or light contribution in our models. Provided the fractional
mass and metals locked up in low mass stars and brown dwarfs in real 
galaxies is small then this is a valid assumption.

\subsection{Gas initial conditions and inflow}
 
We start our models from gas which is initially only very weakly enriched
($Z_o = 0.0002$,  $Y_o = 0.228$ and 
$X_o = 0.7718$, for initial metal, helium and hydrogen gas mass 
fractions respectively). The initial helium mass fraction is from
Pagel et al.\ (1992). The initial metal mass fraction corresponds 
to the lowest metallicity 
data available from W94 SSPs, and it seems reasonable to suppose
that there is some finite start-up time before star formation really 
gets going in a gas cloud, during which a small amount of enrichment occurs. 
Some exploration of varying the initial input parameters 
$Z_o$,$Y_o$,$X_o$ values is described in Section 5.

We allow for enriched (RICH=Y) or un-enriched (RICH=N) inflow,
to simulate the effects of gas inflow known to occur during
interactions and mergers between galaxies (Barnes \& Hernquist 1991). 
In the case of enriched gas we take the enrichment to be the same as
in the existing gas, as if inflow is occurring from neighbouring
regions with similar SFH to the region being modelled.
Inflow of stars formed at earlier times into the region being modelled is
not included, only inflow of gas. 
The inflow rate is allowed to change in our models at specified times 
to simulate e.g. initial collapse, gas inflow during an interaction or 
merger and cessation of inflow. Thus we can explore a range of
possible galaxy histories with these models and test if they can lead 
to observed absorption line-strengths in galaxies.

\subsection{Star formation history}
 
We assume the star formation rate (SFR) is proportional to the gas 
density $SFR=C\rho^{\alpha}$ (M$_{\odot}$ Gyr$^{-1}$), where $\alpha=1$ is 
assumed in our models
and $C$ can be varied. From observations $\alpha$ is thought to lie 
somewhere between 1 and 2 (e.g. Kennicutt 1989). We can allow for a 
possible critical density (RCRIT) below which no star formation takes 
place. However this is set to zero in our models since such a sharp 
cut-off is not well determined observationally (Kennicutt 1989) and does 
not produce very physically realistic behaviour in the models since it 
forces the gas density to sit around RCRIT making star formation switch 
on and off to keep it so. Therefore until we know more quantitative
information about a critical density for star formation it is not 
a meaningful parameter to try to use in these models.
The SFR constant $C$ is related to the SF efficiency in a
closed system (after a short infall period). 
We make use of the timescales from figure 2 of FG94
to identify particular SF efficiencies with different Hubble types 
of galaxies. Table 3 gives the corresponding values for different 
galaxy types.

\begin{table*}
 \centering
 \begin{minipage}{140mm}
  \caption{\bf Star formation efficiencies in different galaxy types.}
  \begin{tabular}{lrcl}
   Galaxy  & Timescale           & SFR Constant  & Reference \\
   Type    & ($\tau_{1/2}$) Gyr & $C$ (M$_{\odot}$ Gyr$^{-1}$)    &  \\
    Sd     & 16  & 0.04  & FG94 \\
    Sc     & 10  & 0.07  & FG94 \\
    Sb     &  3  & 0.23  & FG94 \\
    Sa     &  2  & 0.35  & FG94 \\
    E      &  1  & 0.69  & FG94; Silk 86 \\
   ul\_ir   & SFR $> 10\times$ Spiral & $> 1.0$  & FG94\\ 
\end{tabular}
\end{minipage}
\end{table*}

This completes all the input stellar and gaseous data. Next we describe
how the model is built up. 

\bigskip

\section{Model input, methods and output}

\subsection{Inputs}

A text file is used to specify values for all the model parameters 
we can vary on input. Table 1 shows these parameters for two models
used to generate the chemical evolution and line-strengths in:\\

A) a 'primordial' elliptical (closed system, rapid SF, \\
\indent\indent no gas inflow after a short initial phase of inflow). \\
\indent B) a 'merger' or starburst (modelled by a delayed \\
\indent\indent starburst with gas inflow). \\

The above two models are referred to as reference models in this paper 
and they represent two extremes of galaxy formation. Reference model A 
represents a primordial gas cloud forming stars in a closed system. 
On the other hand, reference model B represents a major merger of two 
gas-rich progenitors, which induces a strong burst of star formation 
via increased SF efficiency and inflow of gas. Inflow of gas is known 
to occur in merging spirals of similar mass (such as the 
'Antennae', NGC 4038/9 Mihos, Bothun \& Richstone 1993, or NGC 7252 
Hibbard \& Mihos 1995). Strong starbursts, detected as ultra-luminous 
infra-red (ul\_ir) sources, occur in galaxies associated with strong 
interactions and mergers (e.g. Sanders et al.\ 1988). Our two reference 
models were chosen as a first test to see if either of these two ideas 
for elliptical galaxy formation (A or B) could predict line-strengths 
comparable with those observed in nearby ellipticals. 

SFR and IMF parameters are included in table 1 as well as 
gas inflow rate, plus changes in SF efficiency and inflow
rate at discrete times after the galaxy first starts to form. Up to two
time changes are included, which can correspond to conditions changing at
the start and end of a later starburst, and/or a period of
increased/decreased inflow. Another input parameter is the time
since the galaxy first started forming (taken to be 17 Gyr -- 
corresponding to the oldest SSP data tabulated by W94).

Using solar metallicity stars FG94 modelled
starbursts due to equal mass, merging spirals. They showed that the
shape of a starburst with time has little effect on the observable
properties soon after the end of the burst. So we do not worry about 
the starburst shape in this work. FG94 found that: pre-burst SFR, 
onset time and duration
of a burst, and average SFR in the burst were the most important
parameters determining the observable colours of their models.

In this Section (3) we concentrate on results from
the two reference model, then in Section 4 a more thorough exploration
of parameter values is carried out to try to obtain the best fits to
line-strength indices observed in elliptical galaxies.
We compare our results with
observed spectral feature strength in early-type galaxies.

\begin{figure*}
\begin{minipage}{140mm}
    \leavevmode
      \epsfig{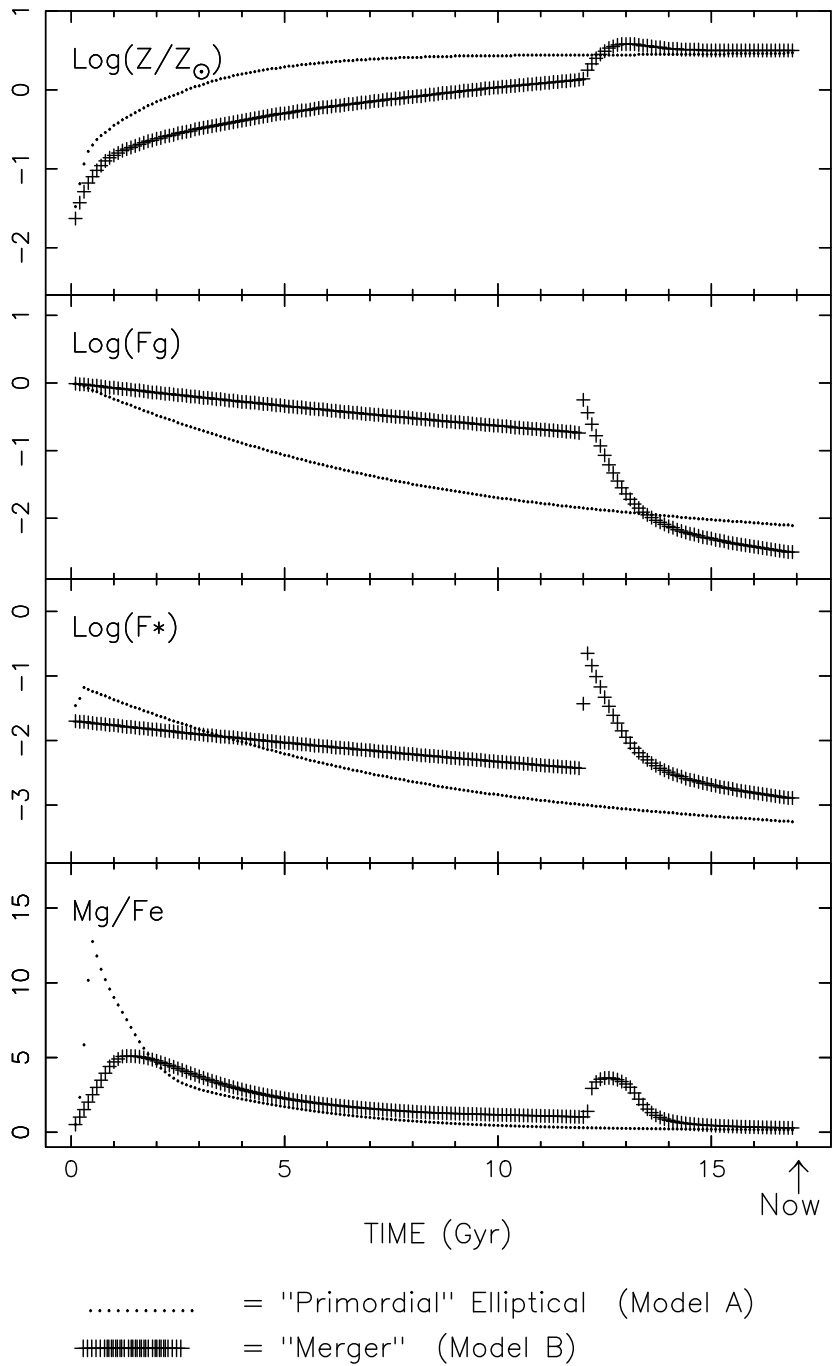}    
\caption{Time evolutions of (A) 'primordial' elliptical model (line of dots) 
and (B) Sbc-Sbc 'merger' model (line of crosses) showing from the top: 
a) ISM metallicity, b) gas mass fraction, c) mass fraction going into stars 
in that time step and d) ISM Mg/Fe mass ratio (for comparison 
(Mg/Fe)$_{\odot}$ = 0.35 from Anders \& Grevesse 1989). 
Mass fractions are normalized by the total mass in the zone 
in each time step.}
    \label{fig:Tevo}
\end{minipage}
\end{figure*}

\subsection{Method}

We start the model with nearly primordial gas. In time steps of $10^8$
years ($>$ SNII lifetimes) stars are then created in the model according 
to the current SFR
and distributed into mass bins according to the IMF. Then we let all the
SNIIs evolve in that time step so that at the end of that step
all the enrichment from those newly formed massive stars is accounted for.
At the end of each step delayed enrichments from SNIa and IMS from
previous time steps are 
accounted for in the ISM. Finally the ISM conditions are updated for any
gas flow which occurred in that step. The resulting ISM conditions 
at the end of the time step are then fed into the next time step.
A check for mass conservation is made at each step by comparing the
expected mass, from the starting mass plus inflow up to that time, to 
the total mass present in stars and gas. These numbers are equal when
the mass loss from stars, SN and inflow from gas is correctly accounted
for.

At each step the initial mass going into stars and the metallicity
of the ISM from which they were made are stored in arrays. The array of
stellar masses for each step is updated, at each step, for mass 
loss by SNIa and IMS from previous steps. These evolved masses and 
metallicities with time are what is required to determine appropriate SSPs
and their weightings, which sum up to make the total galaxy
light. We assume the galaxies are optically thin and use
mass-to-light ratios of the SSPs to evaluate appropriate luminosity
weightings for each step. The SSPs from W94 started with a mass of $10^6$
M$_{\odot}$ integrated over stars with initial masses in the range 
0.1 to 2 M$_{\odot}$. We normalise the W94 SSP luminosities by the ratio 
of initial mass in this range for our data to that in Worthey's SSPs.
This takes account of the mass of the SSP made in each time step in
our models. Since we are looking mostly at features around
5000 \AA, \, we use the V-band luminosity ($L_V$) and mass-to-light ratio
($M/L_V$) to obtain our luminosity weightings.

\subsection{Outputs}

Fig.~1 shows the results of the two reference models
'primordial' and 'merger' (or starburst). The latter is a model with 
a starburst, starting 5 Gyr ago. This figure illustrates the 
time evolution of the models. The primordial model has a burst of star 
formation starting 17 Gyr ago at a rate consistent with the timescale 
given by FG94 for an elliptical galaxy (which is also similar to the 
timescale from Silk 1986). If star formation is stopped after $<2$ Gyr, 
as depicted in Silk (1986), then very low metallicities and line-strengths 
result. Allowing for some inflow in the first few 100 million years 
(as might be expected during the initial galaxy formation) can increase 
the metallicity slightly. In the primordial model depicted in Fig.~1 star 
formation was allowed to continue up to recent times (still in proportion 
to the gas density) to allow the metallicity 
to increase. Even allowing for later, more metal rich stars this initial 
burst (primordial) model {\it cannot} produce the strong absorption lines 
observed in early-type galaxies. This is shown in Fig.~2 where 
line-strengths from elliptical galaxies are compared with these two reference
models. The data shown are from Fisher, Franx \& Illingworth (1995 -- 
hereafter FFI95) and DSP93. 

The delayed starburst model (mimicking a later merger of two gas-rich galaxies)
is able to produce stronger lines since the main light is from stars made
from more metal rich material. The delayed burst model shown in Fig.~1
and Fig.~2 starts 17 Gyr ago with two identical Sbc spirals (in terms of their 
pre-burst SFR). These spirals merge after 12 Gyr, with a SFR typical of 
ul\_ir starburst galaxies, and
a factor two increase in total mass from in-flowing gas, enriched to the
same extent as the existing ISM.

\begin{figure*}
\begin{minipage}{140mm}
    \leavevmode
      \epsfig{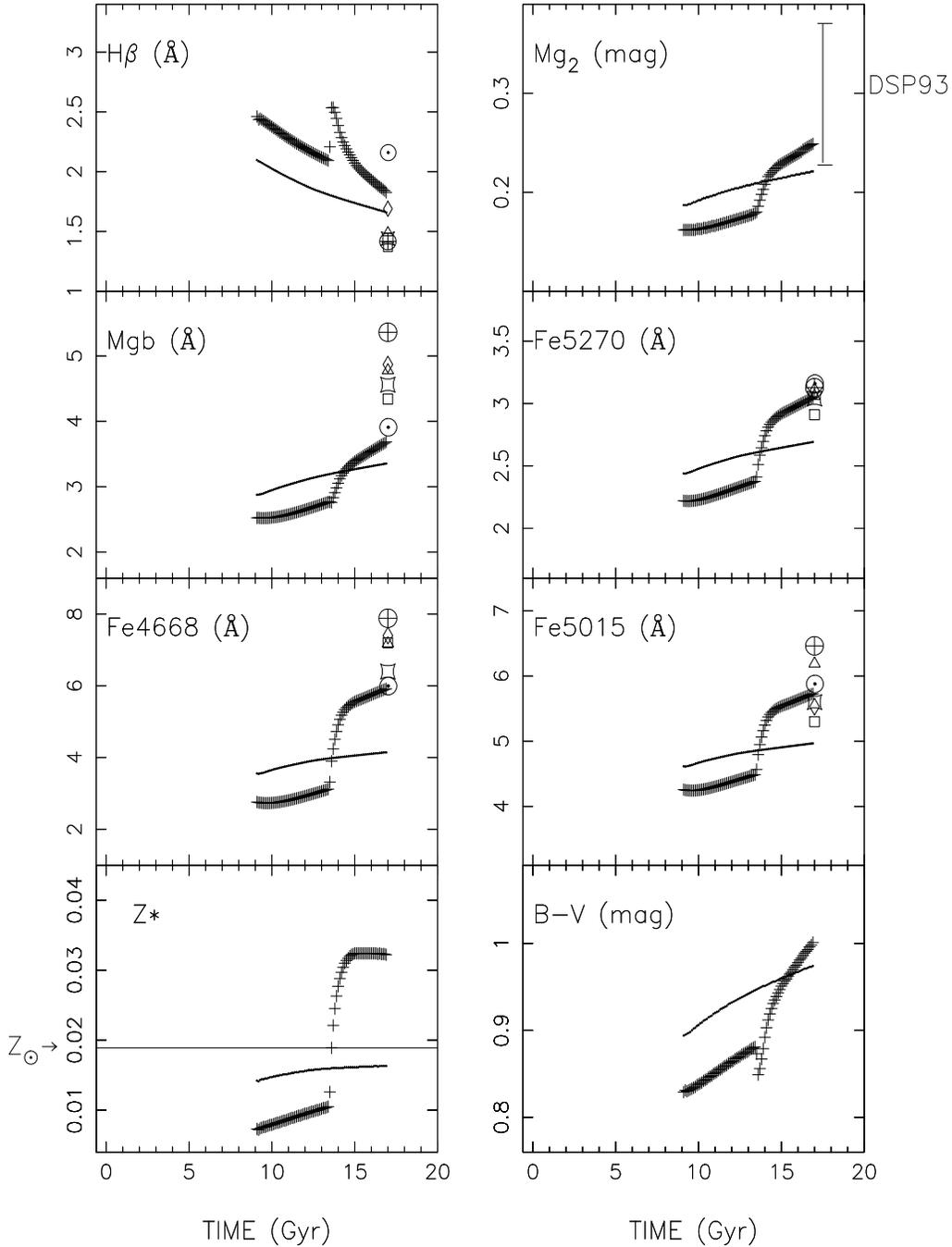}
 \caption{Time evolutions of 'primordial' elliptical model (thin line
of dots) and Sbc-Sbc merger model (line of crosses) showing 
changes in spectral feature strengths from the composite stellar 
population. Also shown are cumulative changes in B-V colour and in 
luminosity weighted metallicity (Z$*$) of the composite population. 
No young, metal
poor stars are included in the SSPs tabulated by W94, therefore
no spectral feature strengths are shown for early time ($<8$ Gyr).
In addition no stars younger that 1.5 Gyr are included in these models
since again these are not modelled in W94. Predicted line-strengths 
from the primordial model are typically much lower than for real 
ellipticals. Line-strengths for 6 bright ellipticals from FFI95 are 
also shown as different
symbols for different galaxies to illustrate how the primordial model
fails to produce strong enough lines. The merger model does better,
however features sensitive to $\alpha$ processes (e.g. Mg and C for 
Fe4668, and Mg for the Mg$b$ index) are still underestimated when Fe
sensitive features can be reached. The range of typical values for the
Mg$_2$ index observed in ellipticals is shown as a vertical bar, estimated
from DSP93.}
    \label{fig:Indices}
\end{minipage}
\end{figure*}

Fig.~2 shows the time 
evolution of spectral features and other stellar parameters in these 
two reference models and illustrates the stronger lines generated in the 
delayed burst (or merger) model (B) compared to the primordial model (A). 
Figs 1 and 2 show examples of the stellar and gaseous output we can 
obtain from our models. The spectral features output ignore any stars 
formed more recently than 1.5 Gyr ago, since the youngest SSP 
in W94 is 1.5 Gyr old. Therefore we note that variations shown in Fig.~2 
at the time of onset of a starburst are unreliable. In addition there 
are no young ($<8$ Gyr), metal poor SSPs in W94, therefore the early 
evolution of feature strengths is not shown here. Later indices are
unaffected by this limitation since we are then seeing old ($>8$ Gyr), 
metal-poor stars and young (but $>1.5$ Gyr), metal-rich stars for which 
SSPs are available. 

Another version of our main program allows the parameter
space of two variables to be stepped through: e.g. starburst time and SFR.
By varying these parameters we can search for the best fit of our models
to the data. This is done in Section 4 for five parameters, fitting 
observed line-strengths in galaxies. 

\subsection{Comparison with other models}

Others have compared the SSPs from various authors (V96;
W94), so we only compare chemical evolution histories here. 
We compared our enrichment model with that of V96.
They have made a complete chemical evolution model, generating their own 
SSPs and following the chemical enrichment of their closed box model
with time from primordial abundances. Time histories of gas metallicity 
and gas mass are compared in Fig.~3. To make this comparison we assumed 
no inflow, no SNIa, a Salpeter IMF and an upper mass limit of 
60 M$_{\odot}$ in order to correspond to the values used by V96.
We assumed the same parameters as V96 for the IMS (case A, $\eta=1/3$ 
and $\alpha=0$ from Renzini \& Voli 1981).
Three different SF efficiencies were compared with those plotted by
V96, their figure 11. These three SF efficiencies are
characterised by $C$=1.92, 0.192 and 0.0192 per Gyr in the SFR equation, 
corresponding to $\nu=19.2$,1.92 and 0.192 per 10 Gyr in the terminology 
of V96).
$C$=0.192 is representative of the solar neighbourhood (Arimoto \& Yoshii
1986). These input criteria were employed for this comparison in order to 
correspond as closely as possible to those used by V96. 
Results from our models are shown at time steps of 0.1 Gyr in Fig.~3. 
Some results from the chemical evolution model of V96 are 
indicated by large open circles at a handful of times through the 
evolution history (estimated from their figure 11). We show two cases from
our models in order also to compare predictions using the M92 and WW 
massive star models: \\

i) FLOSSLIM=0.0 This means we are using only M92 \\
\indent models for SNIIs.

ii) FLOSSLIM=1.0 Means we are using WW data for \\
\indent massive stars up to 40 M$_{\odot}$ and M92 for more massive \\
\indent stars. \\

From Fig.~3 the WW and M92 results from our models give similar 
results, with data from the WW models producing metals slightly 
more rapidly than those from M92. This is as expected since mass loss 
prior to explosion in the M92 models reduces the amount of metals 
which can be generated (particularly in metal-rich stars). 
Also, the conditions for black hole formation are not the same from 
the two sources of SNII models. 

\begin{figure*}
\begin{minipage}{140mm}
    \leavevmode
      \epsfig{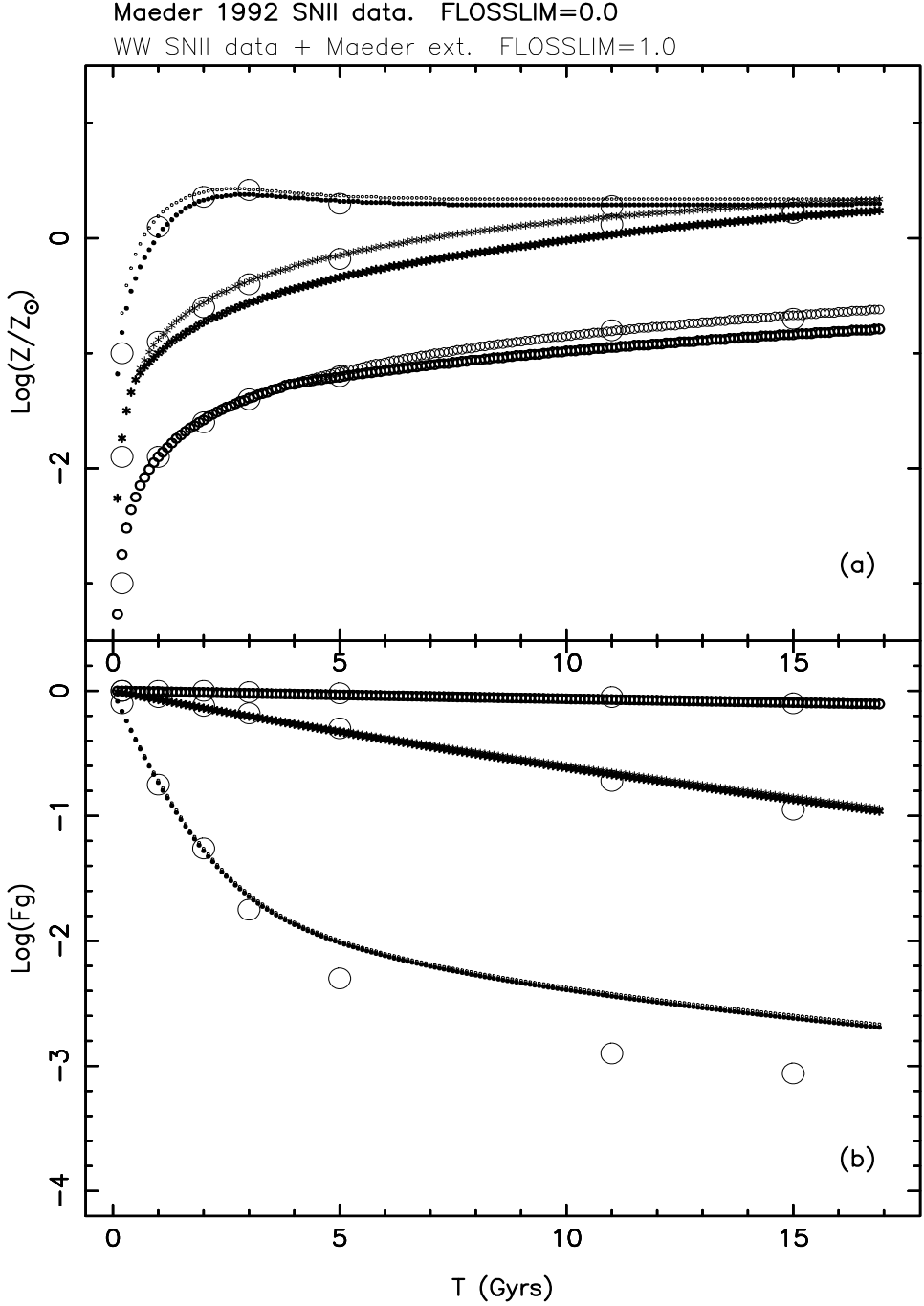}    
\caption{Comparison of models: - Time evolutions of: a) ISM metallicity 
(relative to solar Z$_{\odot}=0.0189$ -- Anders \& Grevesse 1989) 
and b) gas fraction (relative to initial gas 
mass) for models with three different SF efficiencies (C=1.92,0.192 and
0.0192 per Gyr). Large open circles are from the models of V96 
(estimated from their figure 11). Two predictions from 
the current models are shown: 
i) allowing for mass loss (thick symbols, M92 SNII models) and ii) 
for little mass loss (thin symbols, WW SNII models with M92 
extension for stars with initial masses $>40$ M$_{\odot}$).} 
    \label{fig:Zfcomp}
\end{minipage}
\end{figure*}

The values estimated from V96 are similar to our chemical 
evolution model predictions. The main differences appear to be at 
low gas fractions when their models predict less gas left than 
our models. This difference is partly due to different assumptions about
the low mass cut-off (we assume 0.1 M$_{\odot}$, whereas V96 assumed 0.05
M$_{\odot}$) as well as possible differences in the black hole mass assumed
(we assume case C from M92).
However, since the gas content is low ($<1$ per cent of the original gas mass) 
when the differences are large, it has 
little effect on the integrated light and so is unimportant for predicting 
accurate absorption line-strengths. We also note that the full predictions 
of V96 (their figure 11) show numerical oscillations at low gas 
fractions which we do not have in our models. This difference probably 
arises from the different ways we have evaluated the evolution with time: 
we used discrete, coarse time steps (0.1 Gyr) whereas they integrated 
equations. Our method appears more stable against numerical fluctuations 
but produces models which cannot consider systems with very young 
($<$ 0.1 Gyr) starbursts. This latter restriction does not prevent us 
from studying the SF histories of most early-type systems (including 
elliptical galaxies and spiral bulges) which is the purpose of our models.
Our models agree well with the model predictions of metallicity and gas 
fractions by Arimoto \& Yoshii (1986), including at low gas fractions
(C=1.92 in Fig.~3b).

\section{Fits to data}

In Section 3 we found that models of primordial ellipticals with a
single IMF do not produce strong spectral lines. In this Section we
make a more thorough search of parameter space to find out if our 
model with a delayed starburst can describe the observed 
line-strengths for any early-type galaxies. In fact our search will 
include parameter spaces which cover both merger and primordial models.
We use data from FFI95 for ellipticals and brightest cluster galaxies 
(BCGs) on which to try our models. We use their data because they 
include the metallicity sensitive features (Fe4668 and Fe5015) 
as well as the age sensitive feature (H$\beta$), thus helping to break 
the age-metallicity degeneracy. Their features are corrected to the
Lick/IDS standard system (Worthey et al.\ 1994) and corrected for 
velocity dispersion broadening.
Therefore they should be directly comparable with predictions of 
line-strength in the Lick/IDS system (as used by W94 to generate SSPs).
FFI95 give indices within half the effective radius for each galaxy
(i.e. within $r_e/2$).
For chi-squared ($\chi^2$) fitting of models to observed line-strengths 
we also needed estimates of errors for each of the indices used in the
fitting. We obtained estimates of errors using the tables
and plots given in the appendix to FFI95. These errors were taken from
the next to outermost point tabulated, to be conservative, and
these errors, together with the indices used, are given in table 4.
We did not include Fe5335 since the velocity dispersion corrections are 
large for this feature, it was not measured for the BCGs and it is not 
a particularly sensitive feature for discriminating between age and 
metallicity variations. Also shown in table 4 are the galaxy types and
their environments (from Fisher, Illingworth \& Franx 1995).

Five indices are fitted per galaxy. These include: two which are
sensitive mainly to Fe-peak elements (Fe5015 and Fe5270); age
sensitive H$\beta$; and two features which are sensitive to lighter
elements such as C and Mg produced by triple $\alpha$ and $\alpha$ 
capture processes (Fe4668 and Mg$b$). Fe4668 is a strong, metal 
sensitive feature which includes absorption due to Fe-peak elements, 
but also Mg (FFI95) and C; in fact
Tripicco \& Bell (1995) refer to this feature as C4668 because of 
its strong dependence on carbon for cool stars.

To find the best fitting values for model parameters, we stepped 
through about 10 to 20 values in each parameter, starting from zero in
all parameters except the pre-burst and burst SFR constants. Zero for
either of these SFR constants would give us no stars at the start, 
which is just a younger galaxy. Our reference models 
indicated that modelling younger galaxies only aggravates the 
problem of producing strong enough lines compared to the data.
Therefore we have not fitted this possibility, although the fits 
do allow for the possibility of low SFR early on ($C=0.1$ Gyr$^{-1}$). 
The parameters searched through and the steps used are: \\

\noindent 1) Burst onset time (t$*$) in Gyr from the start of formation. 
(0 to 15 in steps of 1 Gyr). \\ 
2) Burst SFR efficiency constant ($C$) in Gyr$^{-1}$. (0.2 to 4.0
in steps of 0.2 Gyr$^{-1}$). \\
3) Inflow rate of gas starting at t$*$ in M$_{\odot}$ Gyr$^{-1}$. (0 to
$10^8$ in steps of $5\times 10^6$ M$_{\odot}$ Gyr$^{-1}$). \\
4) Inflow duration from t$*$ in Gyr. (0 to 0.9 in steps of 0.1
Gyr). \\
5) Pre-burst SFR efficiency constant ($C$) in Gyr$^{-1}$. (0.1 to
0.8 in steps of 0.1 Grys$^{-1}$). \\

The star formation is allowed to continue after inflow has ceased in these
models, but for large values of $C$ most of the gas gets used up rapidly
as shown in Fig.~1c and Fig.~3b.

\begin{table*}
 \centering
 \begin{minipage}{140mm}
  \caption{Line-strength within $r_e/2$ (from FFI95) and error estimates
used for the current model fitting. The errors for Fe5270 were not given
in FFI95 for the BCGs, so are estimated to be the same as those for Fe5015
below. (* indicates a feature is affected by emission, from FFI95.)}
  \begin{tabular}{lllllll}
   Galaxy  & Type and & Fe4668  $\pm$ & H$\beta$ $\pm$ & Fe5015 $\pm$ &
Mg$b$ $\pm$ 
& Fe5270 $\pm$ \\
           & Environment & (\AA) & (\AA) & (\AA) & (\AA) & (\AA) \\
 &&&&&& \\
 \multicolumn{7}{l}{Compact Elliptical (M32)} \\
NGC 221 & cE in local group  &4.82\,\, 0.15 &2.27\,\, 0.07  &5.37\,\, 0.15
&2.93\,\, 0.08 &3.00\,\, 0.10 \\
 &&&&&& \\
 \multicolumn{7}{l}{Ellipticals } \\
NGC 5831 & E in NGC 5813 group  &5.96\,\, 0.33 &2.15* 0.14   &5.86* 0.30   
&3.89\,\, 0.13 &3.15\,\, 0.15 \\
NGC 5846 & E in Virgo Libra group  &6.40\,\, 0.25 &1.44* 0.14   &5.60* 0.37   
&4.56\,\, 0.15 &3.04\,\, 0.19 \\
NGC 2778 & E in Leo cloud group  &7.20\,\, 0.14 &1.37* 0.08   &5.30* 0.21   
&4.34\,\, 0.10 &2.91\,\, 0.10 \\
NGC 7619 & E in Pegasus group  &7.40\,\, 0.16 &1.69\,\, 0.08  &5.53\,\, 0.23  
&4.87\,\, 0.11 &3.10\,\, 0.13 \\
NGC 4472 & E in Virgo S' group   &7.18\,\, 0.08 &1.49\,\, 0.04  &6.19\,\, 
0.11 &4.78\,\, 0.06 &3.06\,\, 0.06 \\
NGC 4649 & E in Virgo S group  &7.86\,\, 0.18 &1.41\,\, 0.08  &6.45\,\, 0.25  
&5.35\,\, 0.12 &3.12\,\, 0.06 \\
 &&&&&& \\
 \multicolumn{7}{l}{Brightest Cluster Galaxies } \\
NGC 2329 & E in poor cluster A 569 &6.57\,\, 0.32 &1.92* 0.13   &4.91* 0.29   
&3.90\,\, 0.12 &2.75\,\, 0.29 \\
NGC 2832 & D in cluster A 779 &8.10\,\, 0.36 &1.41\,\, 0.13  &5.43\,\, 0.32  
&4.78\,\, 0.15 &2.92\,\, 0.32 \\
NGC 4073 & gE  in poor cluster MKW 4  &7.79\,\, 0.34 &1.34\,\, 0.11
&5.49\,\, 0.29  &4.83\,\, 0.13 &2.64\,\, 0.29 \\
 &&&&&& \\
\end{tabular}
\end{minipage}
\end{table*}

\subsection{Fits to ellipticals and BCGs}

Fig.~4 shows the $\chi^2$ contours for the parameter spaces
searched for NGC 5846, one of the ellipticals in the sample of FFI95.
Similar fits and parameter spaces were explored for five other bright 
ellipticals, one compact elliptical and three BCGs for which there 
were at least five line-strengths from FFI95. The best fit model 
parameters found are shown in
Fig.~5 (and table 5) for all these galaxies, in the parameter space of: 
pre-burst SFR constant, burst SFR constant, burst onset time, gas inflow rate 
and duration of inflow during the burst. This covers the major parameters 
found by FG94, plus inflow. We searched the parameter spaces in the order
(a) to (d) as shown in Fig.~4, with starting values of ROOC0=0.2,
FLOWRATE1=$10^7$, TCHANGE2-TCHANGE1=0.1 (duration of inflow). When going
to the next parameter space, best fit values from the previous one were
used. Only 2 or 3 iterations were needed for the fits to converge. 
Although not a full
5 parameter search, this proved to be a robust method for finding best
fits. To test the uniqueness of the best fits we tried varying the
starting values. The best fits were found to be very similar when
different starting values were used, in most cases producing best fits to
within a step of the parameter values given in table 5. In particular the 
burst onset times, the pre-burst and burst SFR constants, and the total 
mass of in-flowing gas seem fairly robust to changes in the starting
values. However, as Fig.~4 indicates, the burst SFR constant $C$ is 
not always well constrained and the inflow duration and rate are not
individually well constrained, rather the duration multiplied by the 
rate (i.e. the total mass of inflow) is the more robust parameter. 
The apparently tight constraint on t$*$ shown in Fig.~4a results from
the need to generate the right metallicity.
From the fits to individual galaxies shown in Fig.~5 and table 5 we
find the following results:

\begin{figure*}
\begin{minipage}{140mm}
    \leavevmode
      \epsfig{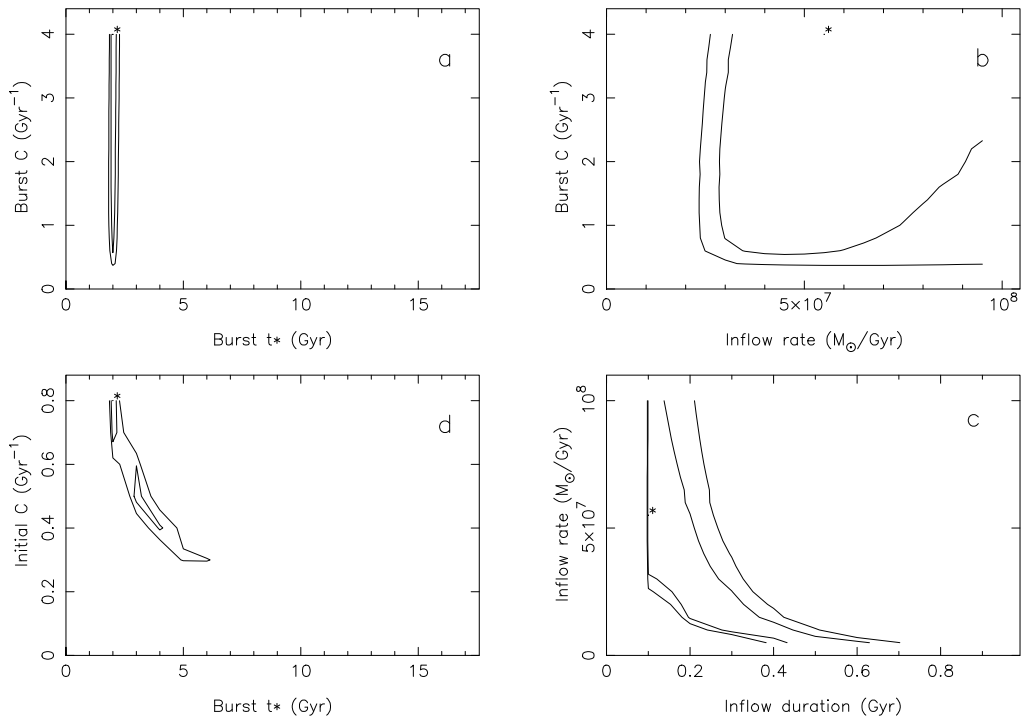}
 \caption{$\chi^2$ contours for parameter spaces
searched for NGC 5846. Contours are at 68 and 90 per cent
confidence levels for two interesting parameters at a time. 
a) Burst SFR constant versus onset time of burst (t$*$). 
b) Burst SFR constant versus inflow
rate at burst. c) Inflow rate versus inflow duration at burst. 
d) Initial (pre-burst) SFR constant versus t$*$.} 
    \label{fig:Chi_n5846}
\end{minipage}
\end{figure*}

\begin{figure*}
\begin{minipage}{140mm}
    \leavevmode
      \epsfig{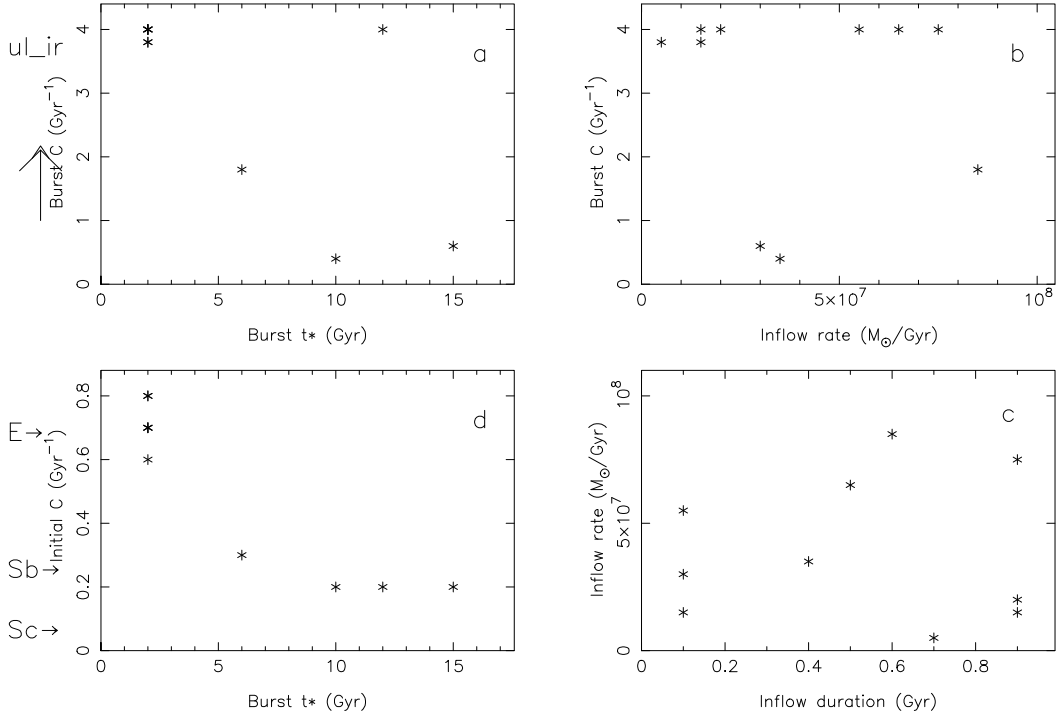}
 \caption{Best fit models found in 4 parameter spaces (as shown in Fig.~4). 
a) Indicates that earlier bursts generally require high efficiencies 
to produce the strong metal lines observed. b) Shows the two parameters 
in our current models which can be used to create high
metallicities. Galaxies therefore cover a wide range in both these
parameters, depending on how strong lined they are. c) For many galaxies
the overall mass of inflow is more important in the models 
than the rate or duration of inflow, so best fits with low inflow 
rates tended to be of longer duration (as the $\chi^2$ contours in
Fig.~4c indicate). d) The parameter space of initial SF efficiencies 
covered those given for
different Hubble types (E to Sc) from FG94. There appear to be two groups 
of best fits with the current models -- one with high initial SF
efficiency followed quickly by a burst and one with lower initial 
SF efficiency (like that in Sb spirals) and a more delayed burst. Typical
SFR for Sc, Sb and E galaxies are indicated in 5d and the range of SFR for 
ul\_ir galaxies is indicated with a vertical arrow in 5a 
(using information given in table 3).}
    \label{fig:Best_fits4}
\end{minipage}
\end{figure*}

\bigskip

  There appear to be two groups 
of best fits with the current models -- one with high initial SF
efficiency followed quickly by a burst and one with lower initial 
SF efficiency (like that in Sb spirals) and a more delayed burst. 
Three ellipticals require late starbursts (NGC 221, 5831 and 2329)
It is interesting to
note that NGC 5831 has a fine structure parameter of $\Sigma$=3.6
(Schweizer \& Seitzer 1992) which also supports the idea that it is a
merger remnant, whereas NGC 5846 has a fine structure parameter
of only $\Sigma$=0.3, and shows no evidence for an intermediate age or 
recent starburst (see t$*$ values in table 5).

\begin{table*}
 \centering
 \begin{minipage}{150mm}
  \caption{Best fit parameters from delayed burst model with inflow.}
  \begin{tabular}{lrcrccc|c}
           &  & \multicolumn{5}{c}{\large\bf Fitted parameters} &
\multicolumn{1}{c}{\large\bf Other parameter} \\
   Galaxy  & $\chi^2_\nu$  & Pre-burst & Burst
& Burst & Inflow rate & Inflow & Starlight \\
           &               & SFR eff. $C$  & onset t$*$
& SFR eff. $C$  & at burst & duration & metals Z$*$ \\
           &               & (Gyr$^{-1}$) & (Gyr) 
& (Gyr$^{-1}$) & (M$_{\odot}$ Gyr$^{-1}$) & (Gyr) & (mass fraction) \\
    &&&&&&& \\
 \multicolumn{5}{l}{Compact Elliptical (M32)} &&& \\
    NGC 221 (cE)  &  2.4 & 0.2 & 15 & 0.6 & 3.0E+7 & 0.1  & 0.027 \\
    &&&&&&& \\
 \multicolumn{5}{l}{Ellipticals} &&& \\
    NGC 5831 (E)  &  0.9 & 0.2 & 10 & 0.4 & 3.5E+7 & 0.4  & 0.031 \\
    NGC 5846 (E)  &  1.7 & 0.8 &  2 & 4.0 & 5.5E+7 & 0.1  & 0.030 \\
    NGC 2778 (E)  &  8.4 & 0.8 &  2 & 3.8 & 5.0E+6 & 0.7  & 0.036 \\
    NGC 7619 (E)  &  9.2 & 0.3 &  6 & 1.8 & 8.5E+7 & 0.6  & 0.039 \\
    NGC 4472 (E)  & 15.7 & 0.7 &  2 & 4.0 & 6.5E+7 & 0.5  & 0.034 \\
    NGC 4649 (E)  & 18.8 & 0.7 &  2 & 3.8 & 1.5E+7 & 0.9  & 0.041 \\
    &&&&&&& \\
 \multicolumn{5}{l}{Brightest Cluster Galaxies} &&& \\
    NGC 2329 (E)  &  3.1 & 0.2 & 12 & 4.0 & 1.5E+7 & 0.1  & 0.035 \\
    NGC 2832 (D)  &  3.7 & 0.7 &  2 & 4.0 & 2.0E+7 & 0.9  & 0.042 \\
    NGC 4073 (gE) &  5.3 & 0.6 &  2 & 4.0 & 7.5E+7 & 0.9  & 0.040 \\
 &&&&&&& \\
\end{tabular}
\end{minipage}
\end{table*}

\begin{figure*}
\begin{minipage}{140mm}
    \leavevmode
      \epsfig{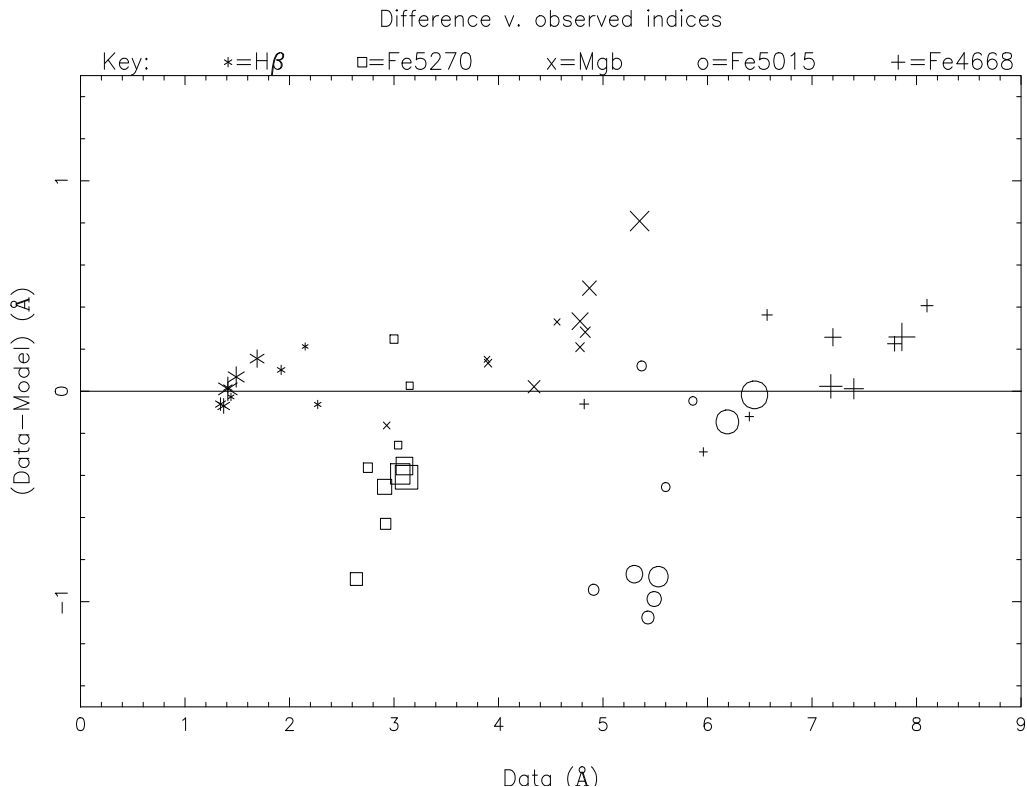}    
 \caption{ Offsets between data and best fit models for 5 spectral
features measured in 10 galaxies. Symbol shapes indicate different 
spectral features. Symbol sizes indicate different galaxies with the 
size increasing for larger reduced chi-squareds of the best fit model.
Smaller symbols tend to lie closer to the model line. The iron sensitive 
lines (Fe5270 and Fe5015) are systematically overestimated by the models 
whereas the lines sensitive to lighter metals (Mg$b$, Fe4668) 
are systematically underestimated. This effect is interpreted as arising
from non-solar ratios and the resultant scatter about models which assume
solar ratios will account for some of the residual differences
between data and models.} 
    \label{fig:Offsets}
\end{minipage}
\end{figure*}

  The model can fit the data in some cases (e.g. NGC 5831, NGC 5846,
NGC 221) but reduced chi-squareds ($\chi^2_\nu$) are still large in most 
other cases. In Fig.~6 we show the offsets of different spectral
features from the best fits to these ten galaxies.
Fig.~6 shows that part of the residual mismatch results 
from non-solar abundance ratios. The comparison of data and model 
line-strengths shows that the observed iron sensitive features 
(Fe5015, Fe5270) are systematically low and observed features which are 
sensitive to lighter ($\alpha$ process) elements (Mg$b$, Fe4668 -- FFI95,
Tripicco \& Bell 1995) are systematically high compared to values from 
the best fit models. Since $\alpha$ process elements and iron-peak elements 
are contributed in different proportions by SNIa and SNII we know that 
this increased light metals to Fe-peak ratio (compared to solar) is telling 
us about the history of SN and SF in these galaxies. 

  Primordial formation (t$*=0$) cannot explain any of these observed 
line-strength data sets in E and BCG galaxies. In order to make stars 
with strong enough lines (particularly metal sensitive lines such as
Fe4668 and Fe5015) some pre-enrichment is required before stars 
which dominate the light are formed.
Vazdekis et al.\ (1997) achieved such enrichment using a 
variable IMF model to obtain strong lines in old stars.
In the current models strong lines can be achieved by one or 
both of the following processes occurring after a {\it delay} (t$*>0$):

 i) {\it inflow} of gas (enriched as in the existing gas), \\
\indent ii) {\it increase in SF} efficiency (caused by a merger or \\
\indent strong interaction). \\

  In the models of star formation in merging galaxies by FG94 solar
metallicities were assumed for the stars throughout. This current work
shows that the bulk of light in ellipticals is from stars with greater
than solar metallicity. This is shown in Fig.~2 where the luminosity
weighted metallicity (Z$*$) of the stellar atmospheres is 1.75 times 
solar for a model which can produce line-strengths at the lower end 
of the range for real ellipticals. Table 5 shows that for the best fitting 
models for individual galaxies Z$*$ lies in the range 0.027 to 0.042, 
which corresponds to 1.4 to 2.2 Z$_{\odot}$.

  The SSPs used in this paper (W94) do not cover the full
metallicity range of the models which best fit the data. The upper limit
of their metallicity is too low for the metallicities required by our
full chemical evolution models. The SSPs go up to about 3 Z$_{\odot}$
whereas we really need up to about 4 Z$_{\odot}$ 
to follow the metallicity of the ISM and stars in our models at later
times (or following a strong burst -- e.g. Fig.~1). Such SSP data are 
now becoming available (Leitherer et al.\ 1996) and we can incorporate 
higher metallicity SSPs into our models in future work. We note that
shallower IMF slopes lead to even higher metallicity (Vazdekis et al.\
1997). In principle X-ray observations of the hot gas ISM in ellipticals 
could tell us about the metallicity of the ISM. However the X-ray
results imply very low metallicities ($\sim 0.4$ Z$_{\odot}$ -- 
Awaki et al.\ 1994), in contradiction with the optical observations 
of strong lines in the starlight. Metallicity ratios from X-ray
observations of the gas will become more constrained using future, 
high spectral resolution X-ray missions such as AXAF and XMM. 

  The nearby, compact elliptical M32 is a useful test of chemical
evolution models. With infrared observations a population of luminous 
asymptotic giant branch stars has been resolved in this companion to 
Andromeda, with an apparent age of $4 \pm 3$ Gyr (Freedman 1992). Our 
delayed burst model (with a burst $\sim 2$ Gyr ago) is consistent with 
this younger age population. 

We further tested if a delayed burst was necessary in all cases by
imagining a model with an initial period of formation with inflow.  
Therefore we tried a model with an extended duration of inflow from 
the start of formation and a single SFR constant throughout, but no
delayed burst. For an inflow duration of 1 Gyr this model could
not produce the required line-strengths. For an inflow duration of 2 Gyr
the lines-strengths could be produced for the cases with old starbursts
(like NGC 4472), but not for cases with younger starbursts (like NGC 5831)
since the H$\beta$ lines in these galaxies are too strong for such old
stars. The SSPs we have used in our models (from W94) do not include
contributions from hot, old stars which may contribute to the strength of
H$\beta$. Poggianti \& Barbaro (1997) recently found that there is some 
sensitivity of Balmer indices to such an evolved star population, but
that H$\beta$ above about 1.5 cannot result from old, metal rich
populations and is dominated by age effects. H$\beta$ is affected by
emission in some cases but this will only decrease the measured index.
Therefore, for observed strong H$\beta$, contamination by emission does 
not remove the need for younger stars in some ellipticals.

  What can we learn about metallicity ratios and SNIa-to-SNII rates
from these fits? We found from Fig.~6 that light metals-to-iron peak 
ratios are generally super-solar confirming previous such results 
(DSP93). However calibrations of line-strengths for 
non-solar ratios are needed to quantify these ratios. 
We do see that those galaxies with weaker lines can be fit
by close to solar ratios (NGC 221, NGC 5831, NGC 5846, NGC 2329 -- tables
4 and 5). The discrepancy of the fits seems to increase with 
line-strength, as the ordering of the ellipticals in tables 4 and 5 shows. 
This is also shown in Fig.~6 where the larger symbols (larger 
$\chi^2_{\nu}$) are generally further from the line where the data 
agree with the model. With SSPs which cover different ratios of
light-to-heavy metals as well as different ages and metallicities we will
be able to constrain the relative proportions of SNIa and SNII
which contribute to the overall enrichment. Fig.~1d shows for example
the drastic change in Mg/Fe ratio in the ISM with time as different
SN types contribute to the enrichment. Reference model A will be
dominated by stars made in the first few Gyr, with very high Mg/Fe ratio,
whereas reference model B will be made of stars with lower Mg/Fe ratio.
Luminosity weighted averages will still exceed solar for model B whilst 
the late burst dominates the light. Fig.~1d indicates that the Mg/Fe
ratio in the ISM might be a useful age indicator since outside of the
bursts the two reference models show similar behaviour of Mg/Fe with 
time.

\section{Discussion}

\subsection{Varying other input parameters}

\subsubsection{Helium-to-metal ratio and black hole formation}

The ratio of helium-to-metal enrichment in the ISM depends on the
SNII explosions and how much mass collapses into black holes.
M92 compares the observed rates of helium-to-metal 
increase ($\Delta Y/\Delta Z \sim$ 3 -- 6) with expected values from 
models with different metallicities and different star mass limits for black
hole formation. In their models the crucial mass is that of the CO core
left at the time of core collapse. In most of our models we have used 
their case C in which any star mass remaining at the time of core collapse
becomes a black hole for CO cores of $>6$ M$_{\odot}$. This produces black
holes of $>22.5$ M$_{\odot}$ from Maeder's massive star models.

In our chemical evolution code we find that the helium-to-metal
increase ($\Delta Y/\Delta Z$) is about 1 to 2. This is lower than 
the observed values in \hbox{H\,{\sc ii}} regions. A steeper IMF leads 
to higher $\Delta Y/\Delta Z$ ratio 
and this ratio is also steeper at low metallicities (M92). M92 find
$\Delta Y/\Delta Z$ = 3.56 and 1.77 for 20th solar and solar metallicity
respectively (for their case C with Salpeter IMF). In most of our
model fits the metallicity is high (apparently higher than solar). 
Therefore it is not so surprising that this ratio is shallow in our 
models. The observed ratio ($\Delta Y/\Delta Z$) from \hbox{H\,{\sc ii}}
regions may not be such a useful constraint to try to match in metal-rich,
early-type galaxies which formed most of their metals early on. The 
observed ratio itself is not well constrained, with estimates ranging 
from about 3 to 6 (Pagel et al.\ 1992). SNIa contribute to the 
metallicity but not the helium, so they act to lower the helium-to-metal 
increase with time. IMS contribute helium and low mass metals. 
More metal rich, massive stars also contribute more helium and metals 
in winds than do metal poor, massive stars. On the other hand, metal rich, 
massive stars do not form black holes. The balance of these processes
leads to a variable $\Delta Y/\Delta Z$ with time and metallicity.
So accurate modelling of the amount and composition of stellar mass loss
as a function of star mass and metallicity is an important requirement
for predicting accurate helium-to-metal enrichment ratios from chemical
evolution models. The strong dependence of $\Delta Y/\Delta Z$ on
theoretical assumptions about SNII explosions limits the usefulness
of $\Delta Y/\Delta Z$ as an observational constraint on chemical
evolution models.

\subsubsection{Pre-enriched gas}

 Can pre-enriched gas produce a good fit to the observations?
We tried a model with {\it no} delayed increase in SF efficiency and 
{\it no} gas inflow, but which is made from pre-enriched gas 17 Gyr ago,
such as might be expected from a very top heavy IMF from which there is
a high level of feedback to enrich the ISM and very 
little material is left trapped in stars. We assumed $C=0.7$ 
(thought to be typical for ellipticals -- FG94, although models
with early galactic winds require much higher values of $C$). 
For those galaxies with very old starbursts (t$*=2$ or less), a
{\it pre-enriched}, single burst model can give comparable fits to 
the data (as those in table 5). The pre-enriched gas required for these
fits starts off with metallicity $Z\sim$ 0.5 to 0.75 solar.
This confirms our earlier suggestion that the stars dominating the light 
at the current epoch have to be made from enriched 
material. However, for other galaxies the observed strength of H$\beta$ 
rules out an entirely old stellar population. These are the galaxies 
which required an intermediate (NGC 5831, NGC 2329) or young (NGC 221)
starburst in our delayed burst models (table 5). 
 
\subsubsection{Models with several starbursts}
 Worthey (1996) suggested that a model of elliptical formation
with several episodes of star formation corresponding to mergers and/or
interactions can fit some of the observational facts (e.g. presence of
kinematic sub-structures, evidence for young stars). Therefore a
future extension to the current models is to try including more
starbursts. However this introduces many more free parameters and the
latest burst will remain important in terms of its relative luminosity.
Worthey (1996) suggested that more bursts at earlier times may explain the 
observations of early-type galaxies. Since we can already produce strong
enough lines with a single delayed burst, more bursts will still not get
round the problem of non-solar ratios, which contributes to the poor
fits of current models to observations of some ellipticals.

\subsubsection{Assumptions about the IMF}
V96 showed that no single IMF, constant in time, can generate the
observed line-strengths in a closed box model with a single SFR constant.
The effects of a changing IMF with time are explored by V96.
They find that a shallow IMF at early times (followed by a normal 
IMF like that inferred for local stars) can produce stellar populations 
with lines as strong as those seen in early-type galaxies. So 'primordial' 
ellipticals with a single IMF are still ruled out for different 
assumptions about the slope of the IMF, but changing the IMF with time, 
from shallow to steep, can produce strong lines. We showed above that 
an early IMF of massive stars (pre-enriching the ISM prior to normal SF) 
can fit the data for some galaxies as well as our merger model. However, 
we note that strong interactions and mergers, accompanied by highly 
increased SF and rapid gas inflow are {\it observed} to occur and that 
both observations and simulations tell us that such events can produce 
early-type galaxies. On the other hand evidence for a variable IMF is 
not so definite (Richer \& Fahlman 1996). In fact Padoan, Nordlund \& 
Jones (1997) recently argued for a universal IMF, arising from the 
statistics of random supersonic flows which simulate conditions during 
star formation. So there is no need for a variable IMF to produce the
observed line-strengths in ellipticals.

\subsubsection{Non-solar abundance ratios}

Non-solar light-to-heavy metal ratios
seem required by the data for most early-type galaxies (e.g. Fig.~6). 
Fits to the observed Mg$_2$ index are improved for models with 
{\it old} stars if the Mg$_2$ calibrations for different Mg/Fe ratios 
from Barbuy (1994) are used. Fig.~7 compares a fit to the data
for NGC 4472, and shows how the Mg$_2$ index is better fitted when
non-solar Mg/Fe is accounted for in these old stars. 

\begin{figure}
    \leavevmode
      \epsfig{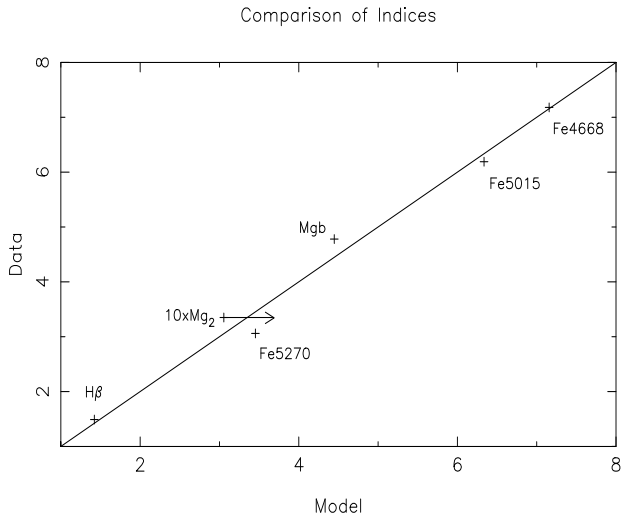}    
 \caption{ Data versus best fit model for NGC 4472. Correction to the
Mg$_2$ index taking into account non-solar Mg/Fe ratios is indicated by a
horizontal arrow. This correction uses the calibration of Barbuy (1994)
to estimate Mg$_2$ indices for old stars with super-solar Mg/Fe, as occurs
in the best fit model for galaxies such as NGC 4472 containing mostly
old, metal-rich stars. Indices below the line are all Fe sensitive 
features. The Mg$_2$ band index is shown $\times 10$ for clarity.}  
    \label{fig:n4472}
\end{figure}

We also tried modelling Mg$_2$ and $<$Fe$>$ (mean of Fe5270 and Fe5335)
features, using SSPs from Weiss, Peletier \& Matteucci (1995) (from the top 
half of 
their table 4). Weiss et al.\ allowed for enhanced $\alpha$-element
compositions and published values for Mg$_2$ and $<$Fe$>$ indices for old
(12 to 18 Gyr), metal rich ($Z\ge$ Z$_{\odot}$) SSPs with Mg/Fe$\ge$
solar. Fig.~8 shows the predictions of some of our models with delayed
starbursts and gas inflow in the Mg$_2$ versus $<$Fe$>$ plane. The rectangular 
area is the region covered by the radial profiles in 9 giant ellipticals 
(data area), from Weiss et al.\ (their figure 8). Models using only solar 
ratio compositions (W94 SSPs) lie outside the region covered by 
ellipticals in Fig.~8.
When we include SSPs with enhanced $\alpha$-element compositions 
we get sequences of models which lie {\it within} the data area.
The models shown in Fig.~8 assume a pre-burst SFR with $C=0.7$ and an 
inflow time of 1 Gyr. Straight lines join models with the same burst
onset time (t$*=0$ or 2 Gyr) but different amounts of inflow (from zero
to 10 times the original mass). Fig.~8 support the idea that 
$\alpha$-element enhanced SSPs in our galaxy models can produce 
feature strengths seen in ellipticals. It will be
interesting to test if this result remains true when other spectral 
features become available (from both observations of ellipticals
and $\alpha$-element enhanced SSP models), particularly when
features over a broader wavelength range are included.

\begin{figure*}
\begin{minipage}{140mm}
    \leavevmode
      \epsfig{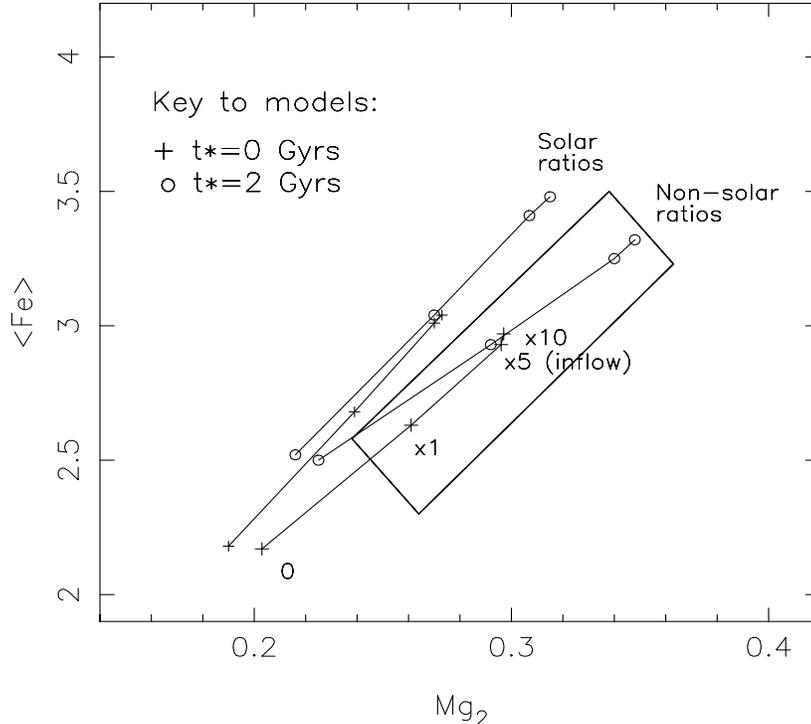}
 \caption{ Predictions from our galactic chemical evolution models 
for two burst onset times and four inflow rates, showing how 
inclusion of $\alpha$-element
enhanced SSPs (from Weiss et al.\ 1995) can produce Mg$_2$ and $<$Fe$>$
feature strengths seen in ellipticals. The rectangular area indicates
where observations of ellipticals lie (from Weiss et al.\ ). Models to the
upper left of the rectangular area assume solar ratios, whereas models
which cover the rectangular area include $\alpha$-element enhanced
compositions.}
    \label{fig:NonS}
\end{minipage}
\end{figure*}

A few cases of the elliptical galaxies from FFI95 can be adequately 
fitted with solar ratios. These tend to be galaxies in which 
there has been substantial late star formation (NGC 221, NGC 5831, NGC
2329, see t$*$ values in table 5). NGC 5846 can also be fairly well 
fitted by the current models with solar ratios. These four cases have 
the lowest metallicity (e.g. weak Fe4668 lines) of all the galaxies 
fitted. From the absolute and relative strengths of the Fe4668 and 
Fe5015 lines in these galaxies, their metallicities (starlight weighted 
average), as well as their metallicity ratios, are 
closer to solar than for the other galaxies. Except for NGC 5846 we 
might say that, qualitatively, their SF histories are more like 
that in the disk of our Galaxy (in that the SF is not restricted to 
early times), than are the SF histories of the other (older) cases
(see t$*$ values).

\subsubsection{Other input parameters}

\noindent {\bf Time step:}
The time step used in these models is assumed to be longer than the longest 
SNII evolution time ($>$ few $\times 10^7$ years for a 10 M$_{\odot}$ 
star). In addition the time step (DT in table 1) must be short enough 
to sample rapid changes, from high SFR or short inflow timescales for 
example ($<$ few tenths of a Gyr). Therefore we tested how the predicted 
line-strengths change when we vary DT over this range for the primordial 
and merger models. Doubling or halving DT from the nominal value of 0.1
Gyr produced variations in line-strengths of $\sim 1/10$th to $1/3$rd
of the error in the observed line-strengths, depending on the feature.
An order of magnitude change in DT produced differences in line-strengths
comparable to the observational errors. Therefore we are confident that
the assumed time step is not strongly biasing the model results, provided
we do not try to model extremely high star formation rates.
   
\noindent {\bf SNIa rate:}
SNIa rates are not accurately known. We have linked the rate used in our
models to
the number of low mass stars present at any time and assume a constant
fraction of those stars are in interacting binaries. If the fraction
of interacting binary stars varies with metallicity or other parameters,
then this assumption will be wrong. We tried changing the SNIa rate between
zero and 10 times the nominal value and running models again for NGC 4472 
and NGC 5831. Zero SNIa rates produced very little change in the minimum 
$\chi_{\nu}^2$ and best fit parameters, and affected mainly the 
helium-to-metal ratio in the gas at late times when the gas mass 
fraction was small. Increasing the SNIa rate by a factor of 10 produced 
a model for NGC 5831 with a slightly earlier burst (t$*=8$ Gyr, $C=0.2$)
and longer inflow ($10^8$ M$_{\odot}$ Gyr$^{-1}$ over 0.9 Gyr), whereas 
for NGC4472 there was little change in the best fit parameters except that the
inflow was less ($5\times10^7$ M$_{\odot}$ Gyr$^{-1}$ over 0.1 Gyr).
Therefore the fits of the models to observed line-strengths do not seem 
to be strongly dependant on the SNIa rate. In particular, the need for
delayed star formation remains. The Mg/Fe and helium-to-metal ratios are
sensitive to the SNIa rate, and we will be able to exploit sensitivities
to element ratios when SSP data are available for non-solar ratios of
light and heavy elements.

\subsubsection{Mass-metallicity relation}

The observed Mg$_2$ versus central velocity dispersion ($\sigma_o$) 
relation for ellipticals (Burstein et al.\ 1988) suggests a close
correlation between the chemical and dynamical evolution. 
If Mg$_2$ scales with metallicity and $\sigma_o$ scales with galaxy mass
this implies a mass-metallicity relation. A weaker correlation of 
Mg$_2$ with galaxy luminosity also re-enforce the idea that there is 
a mass-metallicity relation for early-type galaxies. For the current 
sample of ellipticals the observed Mg$b$ correlates well
with $\sigma_o$ (from McElroy 1995). How do our model fits, with 
line-strengths driven by the SFH, fit this picture? We have not 
attempted dynamical models here, and the data used covers only part 
of each galaxy (along the slit out to $r_e/2$). However
we do find a weak tendency for cases with high infall masses to have
strong lines (correlation coefficient $r\sim 0.5$) and a stronger
tendency for systems with high $\sigma_o$ (observed) to have high
metallicity (Z$*$ from the model fits) (r=0.77).
Therefore we see some evidence that these delayed starburst models with 
infall and a Salpeter IMF may be compatible with dynamical constraints.
Dynamics potentially provide a wide range of additional constraints to the
SFH models. A future aim of this work is to expand the models to
include spatial extent and ensure consistent gas flows between volume
elements. The work of Matteucci \& Gibson (1995) suggests that outflows are
important to eplain abundances in the inter-galactic medium, and may be
needed to explain the mass-metallicity relation. Dynamical constraints 
(e.g. the fundamental plane, gas flows), combined 
with the chemical constraints on models described in this paper, will 
help to clarify the possible SFHs in elliptical galaxies.

\subsection{Future improvements and wish list}

As better SSP data become available (with increased knowledge of stellar
evolution and atmospheric opacities) they can be incorporated into our 
models. In particular, the following improvements are needed to improve 
any chemical evolution models of composite stellar populations:

\begin{enumerate}
\item Higher metallicity SSPs (recently some have become available on 
CD Rom: Leitherer et al.\ 1996, going up to $\sim 5$ Z$_{\odot}$).
\item Improved treatment of later, or missing, stages of stellar evolution
which influence the optical light from stellar populations (current 
limitations are discussed in W94, Timmes et al.\ 1995 and V96). 
\item Calibrations for more indices, particularly age sensitive
features (e.g. H$\gamma$, H$\delta$ lines, Jones \& Worthey 1995).
\item Calibrations for line-strength indices with non-solar element ratios,
particularly for variations in the ratio of light elements -- produced mainly
in SNIIs -- to iron-peak elements -- produced in both SNIIs and SNIas.
Barbuy (1994) has estimated the effect of non-solar ratios on the Mg$_2$ 
index but only at two different Mg/Fe ratios and only for old populations.
Fagotto (1995) and Weiss et al.\ (1995) evaluated some indices 
for SSPs with $\alpha$-enhanced composition and they show these give better 
matches to Mg$_2$ and Fe features than SSPs with solar composition ratios
(see also Section 5.1.5 of this paper).
Such explorations of the effects of non-solar ratios on a range of
spectral features will allow the ratios (Mg/Fe for example) to be used 
as an extra diagnostic for constraining possible SF histories. 
\item We can incorporate SSPs with different IMFs and metallicity ratios
into our models in future to explore the effects of more parameters. Data
with a broader range of spectral features will be needed to fit more
parameters than we have done in this present work. With data covering
more measured features it might be worth investigating other methods for
searching parameter space, such as the Marquardt method used for
example in X-ray spectral fitting packages.
\item Better theoretical understanding of yields from SN models are needed, 
including the amount and composition of mass loss prior to explosions and 
the amount of core mass trapped into black holes.
Accurate modelling of these effects as a function of star mass and 
metallicity is an important requirement
for predicting accurate helium-to-metal enrichment ratios from chemical
evolution models.
\item There are currently many observations of line-strengths in
ellipticals, however, we now have a better understanding of which spectral
features tell us most about the SFH in old galaxies, and which features
are degenerate to age and metallicity changes. Unfortunately many published
results are for these latter types of features which do not distinguish
well between old and metal rich stars. There are also very few
observations of spiral bulges which might have a very different history.
So more observations are needed on both these counts.  
\item Spatially extended models are needed to fully exploit the
gradients in line-strengths observed in early-type galaxies and spiral 
bulges. Eventually this will also aim to make the dynamics of model 
galaxies consistent with their line-strengths.
\end{enumerate}

\section{Conclusions}

We have written a code which can follow the chemical enrichment in 
a region as it evolves due to star formation, evolution and its 
feedback to the ISM. This code allows up to two changes in the SF
efficiency and gas inflow and (using luminosity weighted 
single stellar populations) predicts spectral features on the 
Lick/IDS system (Worthey et al.\ 1994). These predictions are used 
in comparison with observations of composite stellar populations. 
Using this code we have attempted to fit observations of 10
early-type galaxies. 

  We find that ellipticals are not made from primordial gas in a closed zone. 
Some pre-enrichment mechanism is required at early times to produce the strong
lines seen in early-type galaxies. In the current models this enrichment
is achieved by varying the SF efficiency through increased SF rate after
a time delay, and by inflow of gas enriched to the level of the existing
ISM. 

  A delayed starburst (from increased SF efficiency and/or enriched 
gas inflow) can produce the observed strong lines in early-type galaxies,
without the need for a variation in the IMF with time. This simulates 
the effect of a major merger of two massive, gas-rich galaxies. 

  The strength of the H$\beta$ index constrains the age of the delayed 
burst in these models. We found that 7 of the 10 galaxies could be fitted 
with models in which metal rich stars, which dominate the light now, 
formed within the first few Gyr. Assuming a constant IMF
with time, this could be achieved either through a slightly delayed 
burst (t$*$ = 2 or 3 Gyr) following appreciable star formation,
or through an extended period of enriched inflow (over the first $\sim$ 2 
Gyr). The other 3 galaxies required
delayed bursts to explain their line-strengths. These were the compact
elliptical NGC 221 (t$*$=15 Gyr), an elliptical NGC 5831 (t$*$=10 Gyr) 
and a BCG elliptical NGC 2329 (t$*$=12 Gyr), assuming models which
started forming 17 Gyr ago. Therefore there is evidence for later star
formation in some ellipticals in groups and in clusters.

  From our fits to line-strengths measured in ellipticals 
we find that in many cases the $\alpha$-element sensitive spectral 
features are systematically underestimated by the models, whilst iron 
sensitive features are overestimated when SSPs with solar ratios are 
used to model real elliptical galaxies. 
It is interesting that there are
a few exceptions to this general rule (e.g. NGC 221, 5831 and 5846) whose
line-strengths appear to be consistent with solar abundance ratios.
Allowing for non-solar ratios
by using SSPs from Weiss et al.\ (1995), we find that Mg$_2$ and 
$<$Fe$>$ features in ellipticals can be well fitted with our models. 

  Certain spectral features appear more useful than others for constraining 
SF histories in galaxies. These include age indicators such as H$\beta$
and higher Balmer series lines (Jones \& Worthey 1995), plus metallicity 
sensitive features such as Fe5015 and Fe4668 whose ratio changes from 
$<1$ to $>1$ around solar metallicity (W94). Magnesium sensitive features
(e.g. Mg$b$) should also be included for the purpose of exploring Mg/Fe
dependencies. Such features should be put at a high priority in observing 
programmes when studying SF histories in galaxies. 

  In a future paper we will present spectral line-strengths in a sample 
of bulges in early-type spirals (Proctor et al.\ 1997 -- in preparation)
to find out if the current models can fit the spectra of spiral bulges
and whether they also require delayed starbursts to explain their 
line-strengths as do some of the ellipticals we have modelled in this paper.

\section*{Acknowledgments}
The authors wish to thanks David Fisher for helpful
comments on this paper.

\end{document}